**Class Imbalance Corrections Failed to Enhance Discrimination, Model Calibration, and Prediction Stability: An Empirical Simulation Study Based on Clinical Dataset**

Wachiranun Sirikul[1,2,3], Natthanaphop Isaradech[1,4], Wuttipat Kiratipaisarl[1,4], Pakpoom Wongyeekul[2], Noraworn Jirattikanwong[2], Phichayut Phinyo[2,*]

[1]Department of Community Medicine, Faculty of Medicine, Chiang Mai, Thailand 50200

[2]Department of Biomedical Informatics and Clinical Epidemiology, Faculty of Medicine, Chiang Mai, Thailand 50000

[3]Center of Data Analytics and Knowledge Synthesis for Health Care, Faculty of Medicine, Chiang Mai, Thailand 50200

[4]Environmental and Occupational Medicine Excellence Center, Faculty of Medicine, Chiang Mai University, Chiang Mai, Thailand, 50200

**Corresponding Author**

Phichayut Phinyo, MD., Ph.D.

Department of Biomedical Informatics and Clinical Epidemiology, Faculty of Medicine, Chiang Mai, Thailand 50000, **Email:** phichayutphinyo@gmail.com; phichayut.phinyo@cmu.ac.th

## Abstract

**Background:** Class imbalance is a common problem in developing clinical prediction models (CPMs), often leading to poor prediction performance. Various methods for correcting data imbalance have been proposed to improve the development of CPMs. However, the controversy remains whether correcting for imbalance actually improves or harms CPM performances. This study investigated how imbalance correction influenced the classification performances and prediction stability.

**Methods:** This study simulated the development and internal validation of CPM using penalized logistic regression, which employed different imbalance corrections, including algorithm-level rebalance and data-level rebalance by oversampling and combining both over- and under-sampling. The derived dataset for the simulation was the GUSTO-I consisted of 40,830 patients with 2,851 cases. All imbalance corrections were done with various sample size scenarios from 500 to 40,830. Model performance and prediction stability were evaluated using 200 bootstraps to assess model discrimination, calibration, and prediction stability in all levels, including calibration stability, mean absolute prediction error (MAPE), and classification instability indices (CII).

**Results:** All class imbalance corrections did not significantly improve model discrimination. Both data-level and algorithm-level corrections resulted in miscalibration, leading to an overestimation of risk and increased prediction instability, as illustrated by prediction stability, MAPE, and CII plots, compared to the model without the correction.

**Conclusion:** All class imbalance corrections failed to enhance model discrimination and compromised model calibration and prediction stability. Class imbalance is not inherently a pathology that requires corrections. It is not generally advisable to perform class imbalance corrections by default to clinical imbalance data for CPM development.

## 1. Introduction

Clinical prediction models (CPMs) have become essential components of modern medicine, providing clinicians with data-driven support for clinical diagnostic, therapeutic, and prognostic decision-making[1-4]. CPMs often utilize statistical frameworks or machine learning algorithms to learn from patient data and generate predictive estimates for specific health outcomes. Classification is a common task for CPMs, especially in the context of diagnosis. These models provide an estimated probability of a patient having a specific disease or condition and classify them accordingly. The clinical performance of these classification models is fundamentally determined in two dimensions, consisting of discrimination (the ability to accurately distinguish between patients with and without the target outcome) and calibration (the degree of agreement between predicted probabilities and actual observed event probabilities)[5-8].

In the field of medicine, the event of interest—such as sepsis, cardiovascular events, death, or a specific disease diagnosis—occurs infrequently compared to the non-event. This disparity, also known as class imbalance[9], is often characterized by event proportions of less than 20% and even below 1%, which are classified as moderate and extreme imbalances, respectively. Class imbalance poses a significant challenge in clinical classification model development, as models tend to favor the majority class, resulting in a prediction paradox where high overall discrimination performance masks poor performance in predicting the minority class. In the past, the research community considered class imbalance as a pathological defect in data that requires correction, particularly in the machine learning context. The imbalance anxiety has led to the introduction of various imbalance correction techniques, particularly data-level resampling methods[9] (e.g., random oversampling, synthetic minority oversampling technique or SMOTE) and algorithm-level correction. However, these adjustments may unintentionally lead to significant miscalibration of the classification model and possibly introduce bias without enhancing discrimination performance, especially in the context of limited sample sizes[10, 11].

While achieving good discrimination and calibration performance is crucial, recent research highlighted the importance of prediction stability in CPMs[12, 13]. Unstable predictions, susceptible to fluctuations with minor data variations, can lead to inconsistent risk estimations, potentially impacting patient outcomes and eroding trust in these models. This instability can arise from several factors, including small sample sizes, overfitting, and class imbalance.

Therefore, this study aimed to assess the complex relationship between class imbalance corrections, model performance, and prediction stability by simulating CPM development and internal validation employing both data- and algorithm-level imbalance correction techniques across various sample sizes.

## 2. Methods

The empirical simulation study was conducted using an open-source anonymized dataset from the GUSTO Investigators, encompassing 40,830 patient records with 2,851 (6.89%) cases of 30-day deaths [14, 15]. The details on the study design and the eligible criteria of GUSTO-I dataset are described elsewhere[14].

### 2.1. Candidate Predictors

According to a prior study on model prediction stability[13], the following candidate predictors were predetermined: gender, age, elevated blood pressure, hypotension, tachycardia, prior myocardial infarction (MI), ST segment elevation detected on electrocardiogram, and systolic blood pressure.

### 2.2. Class Imbalance Correction

To address the potential class imbalance in 30-day mortality outcomes, four rebalancing strategies were evaluated: (1) using the original imbalanced dataset, (2) implementing algorithm-level rebalancing through class weight adjustments, and (3) and (4) applying data-level rebalancing techniques, specifically oversampling and combined over- and under-sampling. Data-level rebalancing techniques utilized variants of the Synthetic Minority Over-sampling Technique (SMOTE)[16] and Adaptive Synthetic Sampling (ADASYN)[17], which are available in the *imbalanced-learn library* for Python. For oversampling correction, SMOTE for Nominal and Continuous variables (SMOTENC)[16, 18] and ADASYN were selected. For combined over- and under-sampling corrections, SMOTE and Edited Nearest Neighbors (SMOTENN)[19] and SMOTE and Tomek links (SMOTETomek)[20] were chosen. The summary are selected class imbalance corrections is presented in Table A.1. The details method of imbalance corrections is described in the appendix A.

### 2.3. Simulation Study for Model Development and Internal Validation

**[Figure 1. Study Simulation Flow Diagram.]**

The simulation study diagram illustrates model development and internal validation (Figure 1). The primary outcome of interest was 30-day mortality. A penalized logistic regression model was developed using *sklearn.linear_model.LogisticRegression* via Python (version 3.13.9). Model hyperparameters were optimized through grid search in combination with 10-fold cross-validation via *GridsearchCV.* The hyperparameter grids are presented in Table S1. To explore the effect of sample size on model performance and generalizability, the original dataset was randomly down sampled to generate subsets of varying sample sizes and event per predictor (EPP) from 500 (4.375, the minimal required sample[23]), 1000 (8.750),

2000 (17.500), 3000 (26.125), 6000 (34.875), 8000 (52.375), 12000 (104.75), 16000 (139.625), 24000 (209.375), 32000 (279.25) and 40830 (356.375, the full dataset).

### 2.4. Model Assessment

Model performances, optimisms, and prediction stability were evaluated using 200 bootstrap resamples. Discrimination was assessed through the concordance statistics (C-statistic), and the area under the precision and recall curve (AUPRC), along with their optimism. The apparent performance was derived from the training process of each model development approach using its respective training data. The test performance was determined by applying the same model development approach to refit with their bootstrap samples and then validate with the original imbalanced data. The differences between apparent and test performances were defined as optimism. Calibration was examined using calibration plots. To assess stability across predictions[13], the additional metrics were computed and visualized, including discrimination stability, calibration stability plots, mean absolute prediction error (MAPE), and classification instability index (CII). The prediction stability was assessed using the original (apparent) predicted probabilities and 200 predicted probabilities from their bootstrap model (test) predictions on the original imbalanced data. The visualizations for model assessments were created using Seaborn and Matplotlib libraries via Python (version 3.13.9).

### 2.5. Statistical Analysis

The descriptive statistics were performed to explore the derived data qualities and used to present the simulation results as appropriate. To compare the difference between the original imbalanced data and other resampling corrections, we calculate a standardized mean difference (SMD) for each predictor from 200 bootstrap resampling and present it as a mean SMD with 95% CI. A mean SMD above 0.3 indicates moderate differences, while a mean

SMD within the ranges of 0.0–0.1 and 0.1–0.2 indicates non-significant and small differences, respectively. The comparisons for discrimination performances from 200 bootstrap resamples between across the simulation scenarios was conducted using a linear mixed model, adjusted for the full interaction of rebalancing methods and sample sizes. The statistical analyses were performed via STATA version 17 (STATA Corp., College Station, TX, USA) and Python (version 3.13.9).

## 3. Results

### 3.1. The Derived Dataset Characteristics

Of the 40,830 records of patients with acute myocardial infarction (MI) from the derived dataset, the mean age (±SD) was 60.9 (±11.9) years, with a majority being male (74.8%). The prevalences of a history of hypertension and a previous MI were 61.9% and 2.0%. At the baseline, the mean (±SD) systolic blood pressure (SBP) was 128.9 (±23.9) mmHg, and the mean ST segment elevation on the electrocardiogram was 4.1 (±1.9) mm. Additionally, 8.3% of patients presented with hypotension (SBP <100 mmHg), and 32.8% exhibited tachycardia (heart rate >80 bpm).

### 3.2. Class Imbalance Corrections

The summary of mean SMD with 95%CIs from 200 bootstrap samples across sample sizes is illustrated in Figure 2. Age showed mean SMDs consistently above 0.3, which considered moderate differences, while mean SMDs of the other predictors clustered within to 0.0–0.2 range, which considered small and non-significant differences.

**[Figure 2. Comparisons of the Original Imbalanced Data and Rebalanced Data.]**

### 3.3. Model discrimination

Figure 2 shows all discrimination performances across various sample sizes, represented by the median with an interquartile range (IQR) band. For the overall apparent performance trends (Figures 2A and 2B), there was no substantial difference between the original imbalanced data and other class rebalance methods. In contrast, for the test performances and model optimisms (Figures 2C-2F), the original imbalanced data, class-weight rebalanced, and oversampling by SMOTENC showed obviously better performance trends compared to oversampling by ADASYN and combined over- and under-sampling methods. The statistical comparison of rebalanced correction methods versus no correction was conducted using a linear mixed model, adjusted for sample size interaction. The results confirmed that none of the rebalanced methods significantly improved the apparent performances. Furthermore, all data-level corrections exhibited significantly lower test performances and optimism. The detailed information for these statistical comparisons is present in the supplementary materials (Tables S2-7).

**[Figure 3. Discrimination Performances (Median with IQR band) Across Sample Sizes and Methods.]**

### 3.4. Model Calibration

Both data-level rebalancing corrections resulted in model miscalibration, characterized by overestimated predicted probabilities across all sample size scenarios (Figure 4). Although the algorithm-level rebalancing correction induced slight miscalibration at the minimum required samples up to 2000, the model calibrations appeared comparable to those of the other data-level corrections.

**[Figure 4. Calibration Stability Plots Across Sample Sizes and Methods.]**

### 3.5. Prediction Stability Assessment

The prediction stability plots shown in Figure 5 indicate that all class rebalancing methods resulted in a wide range of predicted probabilities, which led to significant prediction instabilities in the models that applied data-level rebalancing corrections. In contrast, the class weight rebalancing exhibited only a modest difference. In contrast, the class weight rebalancing exhibited only a modest difference.

**[Figure 5. Prediction Stability Across Sample Sizes and Method.]**

When considered at the level of individual prediction, the Mean Absolute Prediction Error (MAPE) of individual predicted probabilities of the models employing SMOTENC and SMOTENN were significantly higher than the MAPE from the original imbalanced and class-weight-rebalanced models, as shown in Figure B.1. While MAPE decreased as sample sizes approached the full sample (N=40,830, EPP=356.375), combined over- and under-sampling model (SMOTENN) remained had substantial dispersion of individual predicted probability (MAPE > 0.05).

Furthermore, the assessment using the Classification Instability Indices (CII) plots (Figure B.2) provides a direct measure of classification instability—how often a model changes its decision for an individual based on slightly altered training data by bootstrap resampling. The findings showed that even as the sample size approached the full sample, both data-level rebalancing corrections produced unstable predicted probabilities for individuals around the decision cut-off at 0.10, especially for the models that used combined over- and under-sampling data (SMOTENN). In summary, all derived models that incorporated oversampling and combined resampling corrections exhibited more prediction instability in all sample size scenarios, whereas class weight rebalanced showed marginally differences on the model prediction stabilities. More details experiment results of the prediction stability assessment are presented in Figures S1-4.

## 4. Discussion

Our findings underscore a relationship between class imbalance corrections, model performance, and prediction stability in the development and internal validation of CPMs. While all modelling approaches achieved high C-statistics for 30-day mortality, all class imbalance corrections failed to improve the model's actual ability to classify the minority target class. This was evidenced by consistently low AUPRC across all approaches and sample size scenarios. Furthermore, addressing class imbalance through various techniques did not inherently enhance both overall and positive class discrimination, as well as decrease their optimism. In fact, utilizing complex resampling methods on relatively simple clinical data appeared to degrade model discrimination across all evaluated aspects. Our findings were consistent with the previous studies[10, 11, 24], which caution that imbalance corrections often result in inaccurate probability estimates without offering any tangible benefit in discrimination. The discordance between the C-statistic and AUPRC observed in our study demonstrated that AUPRC was particularly sensitive to the false positives introduced by oversampling. This led to an increase in recall (sensitivity) but resulted in a significant decrease in precision (positive predictive value)[24-26]. Our findings on model calibration assessment also confirmed that correcting class imbalances, especially data-level corrections, resulted in models that exhibited strong miscalibration, characterized by overestimated predicted probabilities across all sample size scenarios. Even calibration plots showed greater stability in larger samples. The models that utilized both algorithm-level and data-level rebalancing corrections still exhibited miscalibration with overestimated probabilities. This finding was consistent with the previous study that raised concerns about utilizing data-level imbalanced corrections for CPM development, which could lead to inaccurate probability estimates and poor decision-making[10, 11, 27].

In this study, we extended beyond traditional metrics by further evaluating prediction stability, which is a newly emerging metric in CPM research[12, 13]. This aspect includes the stability of the C-statistic, as demonstrated by test performance on the original data, calibration instability, and fluctuations in the MAPE and CII. Our study results revealed that all class imbalance corrections, particularly data-level resampling approaches, showed a significant instability in all aspects of prediction stability. The observed instability can be attributed to how rebalancing strategies modified the model's training processes. The primary driver of the observed instability appeared to be coefficient inflation. By artificially increasing the recognition of the minority class, the model is forced to assign more weight to these features, leading to more extreme and widespread predicted probabilities. Data-level resampling, particularly SMOTE variants, introduces synthetic noise that shifts the decision boundary. This creates a trade-off where the model achieves more positive predictions at the cost of a substantial increase in false positives. As shown in the CII results, these inflated coefficients made the model highly sensitive to minor perturbations in the training data, leading to unpredictable classification changes for patients near the decision cut-off. To mitigate the overfitting and coefficient inflation inherent in resampling, we utilized penalized logistic regression as our development framework. This approach serves as a necessary counterweight, attempting to regularize the model and prevent the coefficients from reaching the extreme levels triggered by resampling. Subsequently, it is still insufficient to stabilize models trained on resampled data, as demonstrated in the prediction stability results. In contrast, class-weight rebalanced correction could maintain stability levels that were nearly identical to model training with the original imbalanced dataset, indicating its robustness in comparison to data-level resampling. However, adjustments based on class weights must be applied cautiously to ensure that any trade-off in discrimination resulting from increased model recall (sensitivity) is acceptable, given the potential for more false positives and miscalibration due to overestimated risk predictions.

Currently, the cumulative evidence of data imbalance corrections remains scattered and inconclusive, given the heterogeneity of event fractions, thresholds, and reporting across different diseases and CPM development frameworks. There was one ongoing scoping review on resampling methods for class imbalance in CPMs awaiting to be explored[28]. However, the previous simulation and empirical studies suggested that effective sample size planning, rather than aggressive post-hoc balancing, often negates the need for resampling[12, 13, 27, 29-32]. The review[33] and studies [34, 35] on the algorithm-level correction also showed that cost-sensitive learning directly penalizes errors in the minority class and outperforms methods that operate at the data level in ML context.

The present literature and our findings collectively suggested that for many clinical applications, the optimal strategy may conservatively use the original imbalanced data for CPM development while adjusting the decision threshold to achieve the desired sensitivity and specificity balance, along with external validation for specific clinical utility.

### 4.1. Study Limitations

The first limitation of this study was that it was simulated on a single case study. The model performances and degree of instability may vary across different clinical domains or datasets with varying predictor complexities. However, we selected the GUSTO-I dataset for our study experiment, as its large sample size provided the necessary statistical power to rigorously evaluate prediction stability. Additionally, this dataset had the potential to represent standard CPMs commonly used in real clinical practice by utilizing a parsimonious model with straightforward clinical predictors. Second, this study only examined one event fraction, considering a moderate level of imbalance. It could be argued that the stability of model predictions may be affected differently by varying class imbalance severity and that our results might not apply in other contexts where the severity of the class imbalance varies. However, the recent study [10] examining three different event imbalance fractions (0.01%, 0.10%, and 0.30%) found that the distortion in probability estimates and calibration caused by corrections such as SMOTE was more pronounced for minority classes with a smaller percentage (0.1%). Moreover, employing resampling imbalance corrections at all levels of imbalance did not yield significant benefits in discrimination and resulted in strong miscalibration with overestimated risk predictions. Third, because our models were developed using penalized logistic regression, the study findings may not generalize to the behavior of "black-box" ML algorithms applied to similar class imbalance corrections[36]. However, we selected penalized logistic regression as an understandable, standard framework for CPM development that effectively mitigates overfitting and coefficient inflation through fine-tuned regularization. Furthermore, current systematic reviews indicate no consistent performance advantage of machine learning over logistic regression when the event-per-variable ratio is adequate[37]. Lastly, this study was focused on the development and internal validation processes. Although we rigorously assessed model performance and stability, this study did not provide empirical evidence on how these resampled models would perform in geographically or temporally distinct external validation. However, it could be logically inferred that internal validation performances and stability are a prerequisite for external generalizability[7, 8, 13, 38-40]. If the derived model shows considerable miscalibration and prediction instability during internal validation—when the data distribution is relatively stable—it is improbable that the model framework would maintain its robustness when faced with the case-mix variations and dataset shifts that are common in external validation settings[5, 8, 38-41].

## 5. Conclusion

This study demonstrated that the class imbalance corrections in clinical prediction model development led to suboptimal predictive performance. Apparent discrimination metrics appeared deceptively stable, while test discrimination metrics and their optimism, calibration, and prediction stability at all levels showed significant compromise. Our results indicated that data-level resampling techniques—especially ADASYN, SMOTE, and their variants—generated synthetic noise and coefficient inflation. This led to a systematic overestimation of risk probabilities and increased sensitivity to minority groups in training data. Although algorithm-level correction using class weighting was more robust, it still required a trade-off between enhanced recall and reduced precision. Class imbalance should not be considered an inherent pathology that needs correction. Instead, the integrity of a CPM is maintained by utilizing the original imbalanced datasets alongside optimized decision thresholds and comprehensive sample size planning.

**Acknowledgements**

The study was supported by the Faculty of Medicine, Chiang Mai University and Chiang Mai University.

**Author contributions**

**Wachiranun Sirikul:** Conceptualization, Methodology, Data curation, Formal analysis, Visualization, Investigation, Validation, Writing - Original Draft, Writing - Review & Editing, Software. **Natthanaphop Isaradech:** Conceptualization, Methodology, Data curation, Formal analysis, Visualization, Investigation, Validation, Writing - Original Draft, Writing - Review & Editing, Software. **Wuttipat Kiratipaisarl:** Conceptualization, Methodology, Validation, Writing - Review & Editing. **Pakpoom Wongyeekul:** Conceptualization, Methodology, Validation, Writing - Review & Editing. **Noraworn Jirattikanwong:** Conceptualization, Methodology, Validation, Writing - Review & Editing. **Phichayut**

**Phinyo:** Conceptualization, Methodology, Formal analysis, Visualization, Investigation, Validation, Writing - Original Draft, Writing - Review & Editing, Resources, Supervision, Project administration.

**Ethics approval**

This study received exemption from ethics review by the ethical committee of the Faculty of Medicine at Chiang Mai University (certificate of exemption no. 0298/2568). The exemption was granted because the study involved a secondary analysis of de-identified data from a previously approved study focused on developing a clinical prediction model.

**Funding**

This study receives no external funding.

**Conflict of interest**

None declared.

**Declaration of Generative AI and AI-assisted technologies in the writing process**

During the preparation of this manuscript, the authors used Gemini and QuillBot to assist with language editing, grammar correction, and clarity. All AI-assisted outputs were reviewed, edited, and verified by the authors, who take full responsibility for the content of the manuscript.

## Appendices

### Appendix A: Class Imbalance Correction Strategies

**Table A.1** Summary of Class Imbalance Correction Methods

| Strategy Type | Method | Key Mechanism |
|---|---|---|
| Baseline | Original Dataset | No correction applied. |
| Algorithm-Level | Class Weighting | Modifies the loss function to penalize minority class errors more heavily using inverse frequency weights. |
| Oversampling | SMOTE and SMOTENC | Generate synthetic minority samples by interpolating between existing minority instances and their nearest neighbors. |
| Oversampling | ADASYN | Similar to SMOTE but adaptively generates more data for minority samples that are difficult to learn based on their nearest neighbors. |
| Combined | SMOTENN | Applies SMOTE oversampling followed by Edited Nearest Neighbors (ENN) to remove noisy samples and clean the decision boundary. |
| Combined | SMOTETomek | Applies SMOTE oversampling followed by Tomek Links to remove overlapping instances between classes. |

### A.1 Data-Level Rebalancing: Oversampling

#### Synthetic Minority Over-sampling Technique (SMOTE) and SMOTE for Nominal and Continuous Variables (SMOTENC)

SMOTE[16, 18] is the most common oversampling technique for an imbalance correction in both engineering and medical literature. In contrast to Random Over-sampling (ROS), which simply duplicates existing minority cases, SMOTE creates synthetic samples from the minority class using the information available in the dataset.

For each minority sample $x_i$, SMOTE selects $k$ nearest neighbour from the minority class. It then randomly chooses one of nearest neighbour $\hat{x_i}$ and generates a new sample $x_{new}$ at a random point along the line connecting $x_i$ and $\hat{x_i}$:

$$x_{new} = x_i + u \cdot (\hat{x_i} - x_i) \qquad \text{Eq. (A.1)}$$

, where $u$ is a random number between 0 and 1. $u$ was the same for all variables but varied for each SMOTE sample. This approach ensures that the SMOTE sample is located on the line connecting the two original samples used to create it. SMOTE rebalances the minority distribution by filling the gaps among minority samples, thereby theoretically making the decision boundary more linear and distinct. This assumes that the local area of a minority sample primarily consists of other minority samples. However, SMOTE may generate synthetic data in noisy regions, thus joining the classes rather than separating them. When working with mixed data types, which include both continuous and categorical features, SMOTENC serves as an extension of the SMOTE algorithm that handles categorical data in a distinct manner.

**Adaptive Synthetic Sampling (ADASYN)**

The ADASYN approach[17] uses weighted distribution as a criterion to automatically determine the number of synthetic samples needed for each minority data point. This algorithm focuses on generating more synthetic data for minority class examples that are considered "difficult to learn" and are wrongly classified by $k$ Nearest Neighbors, whereas the basic implementation of SMOTE does not differentiate between easy and difficult-to-learn samples when applying the nearest neighbors' rule. This method enhances learning concerning data distributions in two main ways: (1) it reduces the bias caused by class imbalance, and (2) it adaptively shifts the classification decision boundary toward more difficult-to-learn examples. However, ADASYN can exacerbate overfitting by oversampling difficult-to-learn examples.

### A.2 Combine Over- and Under-Sampling using SMOTE Variants

As previously mentioned, oversampling by SMOTE can generate noisy samples by interpolating new points between marginal outliers and inliers. The concept of combining oversampling with under- sampling can be solved by cleaning the decision space resulting from oversampling. In this regard, Tomek's link (Tomek)[21] and Edited Nearest Neighbors (ENN)[22] are the two under-sampling methods that have been added to the pipeline after applying SMOTE oversampling to obtain cleaner decision space. Both approaches, SMOTE and Edited Nearest Neighbours (SMOTENN)[19] and SMOTE and Tomek links (SMOTETomek)[20] are available via the *sklearn* library. Tomek links under-sampling will identify pairs of instances from opposite classes that are close neighbors and remove them to clean up the decision boundary. For ENN, the motivation behind this method is similar to Tomek. ENN tends to remove more examples than the Tomek links do, so it is expected that it will provide more in-depth data cleaning. ENN is used to remove examples from both classes. Thus, any example that is misclassified by $k$ nearest neighbors is removed from the training set. While these approaches have the potential to improve classification performance, they involve aggressive data modification that distorts the probabilistic structure of the cohort.

### A.3 Algorithm-Level Corrections: Class Weighting

Class weighting (or cost-sensitive learning) modifies the loss function utilized by the classifier, rather than adjusting the data distribution as seen in other data-level corrections. In standard classifiers, the loss function treats all misclassifications equally. In the presence of a significant class imbalance, this often leads the model to prioritize the majority class to minimize the overall error rate. Class weighting counteracts this by assigning a higher penalty (weight) to misclassifications of the minority (positive) class. The loss function is calculated using $y$ as the observed binary label (where 1 represents the positive outcome and 0 represents the negative) and $\hat{y}$ as the model's predicted probability for that observation:

$$Loss = -\sum(w_1 \cdot y \cdot \log(\hat{y}) + w_0 \cdot (1-y) \cdot \log(1-\hat{y})) \quad \text{Eq. (A.2)}$$

where $w_1$ and $w_0$ are the weight assigned to the positive and negative classes, respectively.

In this study, weights were determined during the model hyperparameter tuning via 10-fold *GridsearchCV.* The weight parameters associated with each class are defined in the format {class_label: weight}; for instance, $\{0: w_0,\ 1: w_1\}$ indicates that the penalty weights ($w_i$) assigned to the positive (class_label=1) and negative (class_label=0) classes are represented by the values provided. For the default or "None" class weight adjustment, $w_0$ and $w_1$ are equal to 1. For the "balanced" mode, $w_i$ are automatically determined using the inverse frequency method in the input data:

$$Postive\ class\ weight\ (w_1) = \frac{N}{2 \cdot N_{pos}} \quad \text{Eq. (A.3)}$$

$$Negative\ class\ weight\ (w_0) = \frac{N}{2 \cdot N_{neg}} \quad \text{Eq. (A.4)}$$

, where $N$ is the total number of samples, and $N_{pos}$ and $N_{neg}$ are the counts of the respective classes. The parameter grid for fine-tuning class weight of a penalized logistic regression was {"balanced"} and $\{0: 1,\ 1: w_1 |\ 2 \leq w_1 \leq 20\}$. This approach ensures that the minority (positive) class exerts a proportional influence on the model's parameters during training, preventing the majority (negative) class from dominating the learning process.

## Appendix B: Prediction Stability Results

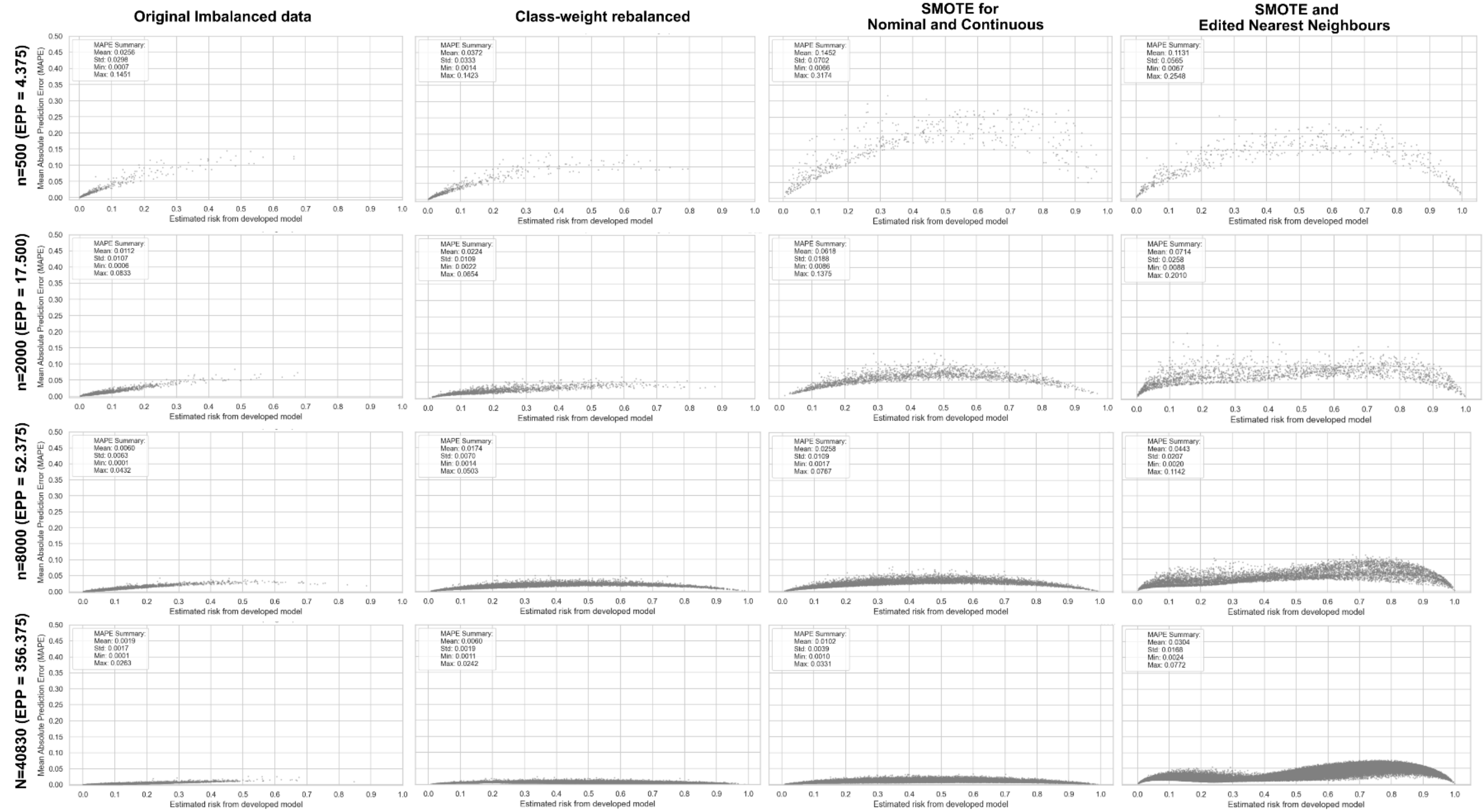

**Figure B.1 Mean Absolute Prediction Error Across Sample Sizes and Methods.** The range of estimated risk from the developed model is plotted on the x-axis against the corresponding Mean Absolute Prediction Error (MAPE) on the y-axis. MAPE for each individual observation is calculated as the average absolute difference between the estimated risk from the developed model and the predicted probabilities generated by 200 bootstrap models using the same modelling approaches. The figure illustrates four methods for addressing class imbalance, including the original imbalanced data as a baseline, class-weight rebalancing, SMOTENC (representative of oversampling approaches), and SMOTENN (representative of combined resampling approaches). Four landmark sample sizes along with their respective EPP values are presented. Within each plot, scatter points represent the absolute prediction errors for individual estimated risks. An inset text box within each panel presents a summary of the MAPE distribution. **Abbreviations:** EPP, Events Per Predictor; MAPE, Mean Absolute Prediction Error; SMOTENC, Synthetic Minority Over-sampling Technique for Nominal and Continuous Variables: SMOTENN, Synthetic Minority Over-sampling Technique and Edited Nearest Neighbors; Std, standard deviation.

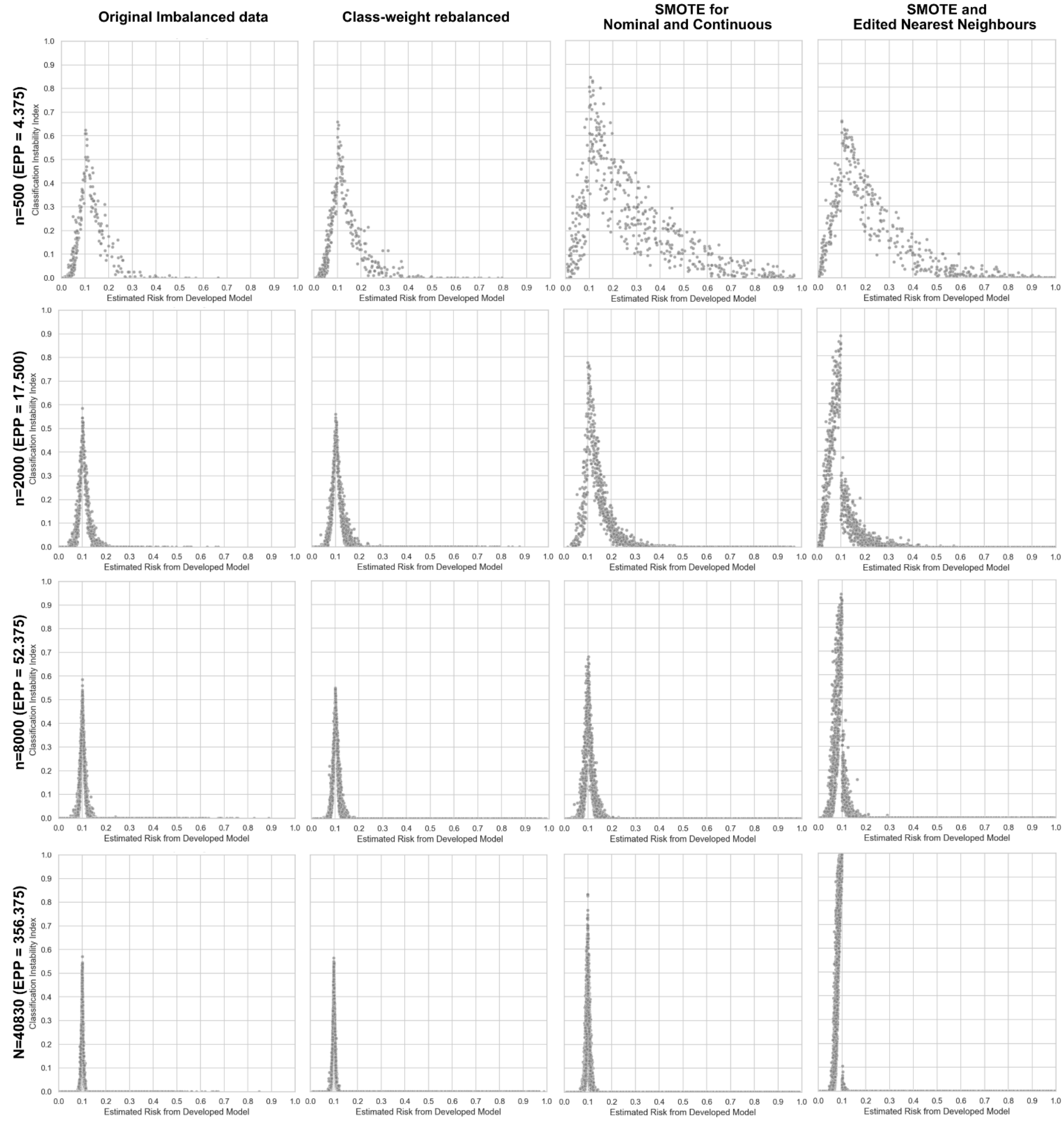

**Figure B.2 Classification Instability Indices Across Sample Sizes and Methods.** For each individual prediction, the CII was calculated by applying a predefined decision cut-off (0.10) to the predicted probabilities generated across the 200 bootstrap models, measuring how frequently the model's binary classification changed due to variations in the bootstrap models using the same modelling approaches. The range of estimated risk from the developed model is plotted on the x-axis against this corresponding CII on the y-axis. The figure displays four approaches to class imbalance, including the original imbalanced data as a baseline, class-weight rebalanced, SMOTENC (representative of oversampling approaches), and SMOTENN (representative of combined resampling approaches). These are presented across four landmark sample sizes along with their respective EPP values. **Abbreviations:** CII, Classification Instability Indices; EPP, Events Per Predictor; MAPE, Mean Absolute Prediction Error; SMOTENC, Synthetic Minority Over-sampling Technique for Nominal and Continuous Variables: SMOTENN, Synthetic Minority Over-sampling Technique and Edited Nearest Neighbors.

## Figures

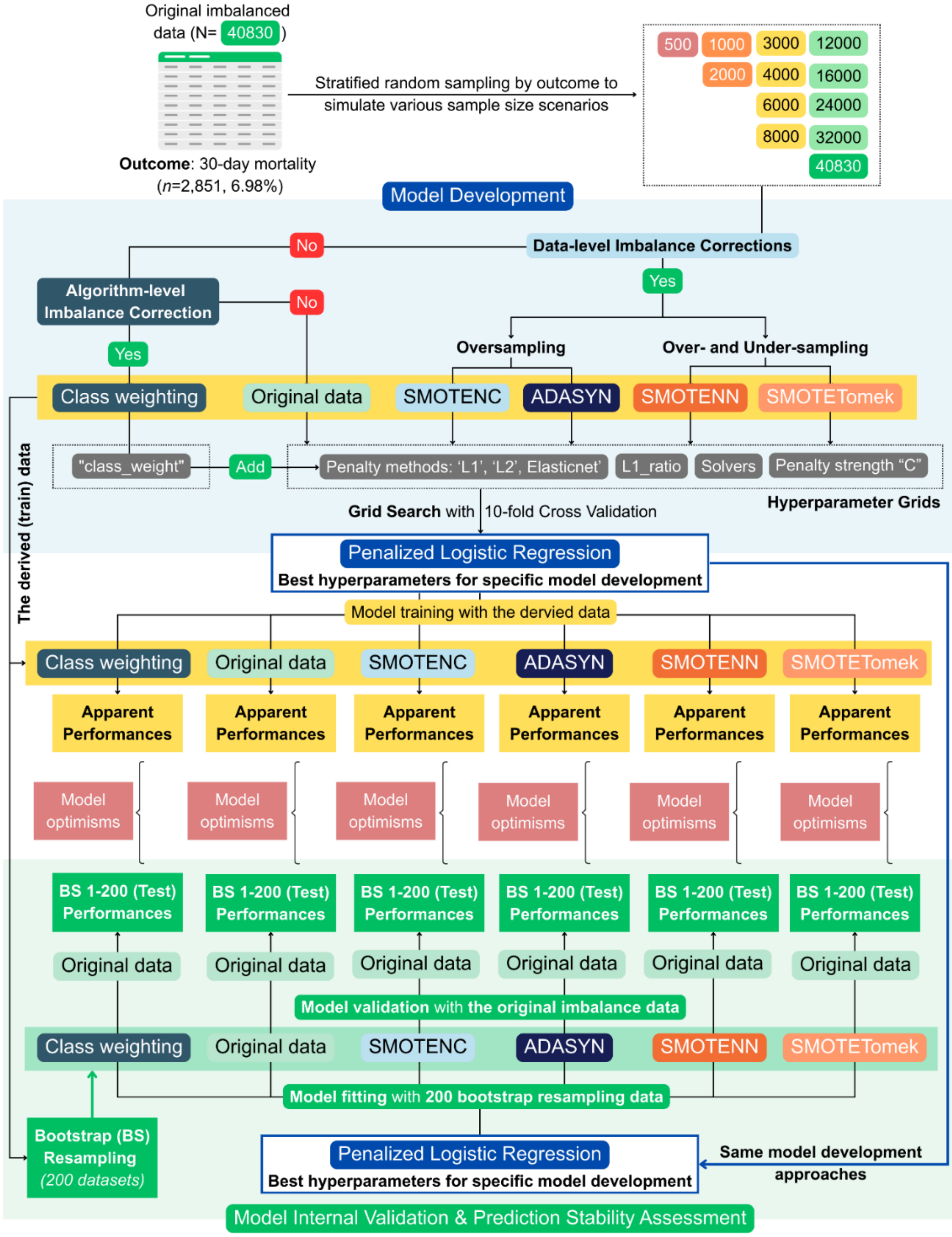

**Figure 1. Study Simulation Flow Diagram.** The empirical simulation started with imbalanced data. Sample sizes from 500 to 40,830 were simulated using stratified random sampling. The model development phase evaluated six class imbalance solutions: (1) using the original data as a baseline, (2) algorithm-level correction via class weighting, (3) oversampling via SMOTENC, (4) oversampling via ADASYN, (5) combined sampling via SMOTENN, and (6) combined sampling via SMOTETomek. Grid search with 10-fold cross-validation was used to train penalized logistic regression models for each pathway to find the best hyperparameters (including L1, L2, and Elastic Net). Apparent performance was training performance with each approach. Finally, model internal validation and prediction stability

were assessed using 200 BS datasets to refit the models with the same development approaches and validate against the imbalanced data to obtain 200 test performances. The difference between apparent and test performances determined model optimism. The prediction stability was assessed using the original (apparent) predicted probabilities and 200 BS test probabilities. **Abbreviations:** ADASYN, Adaptive Synthetic Sampling; BS, Bootstrap; C, inverse of regularization strength; CV, cross-validation; Elasticnet, elastic net regularization; L1, Lasso regularization; L2, Ridge regularization; SMOTE, Synthetic Minority Over-sampling Technique; SMOTENC, SMOTE for Nominal and Continuous variables; SMOTENN, SMOTE and Edited Nearest Neighbors; SMOTETomek, SMOTE and Tomek links.

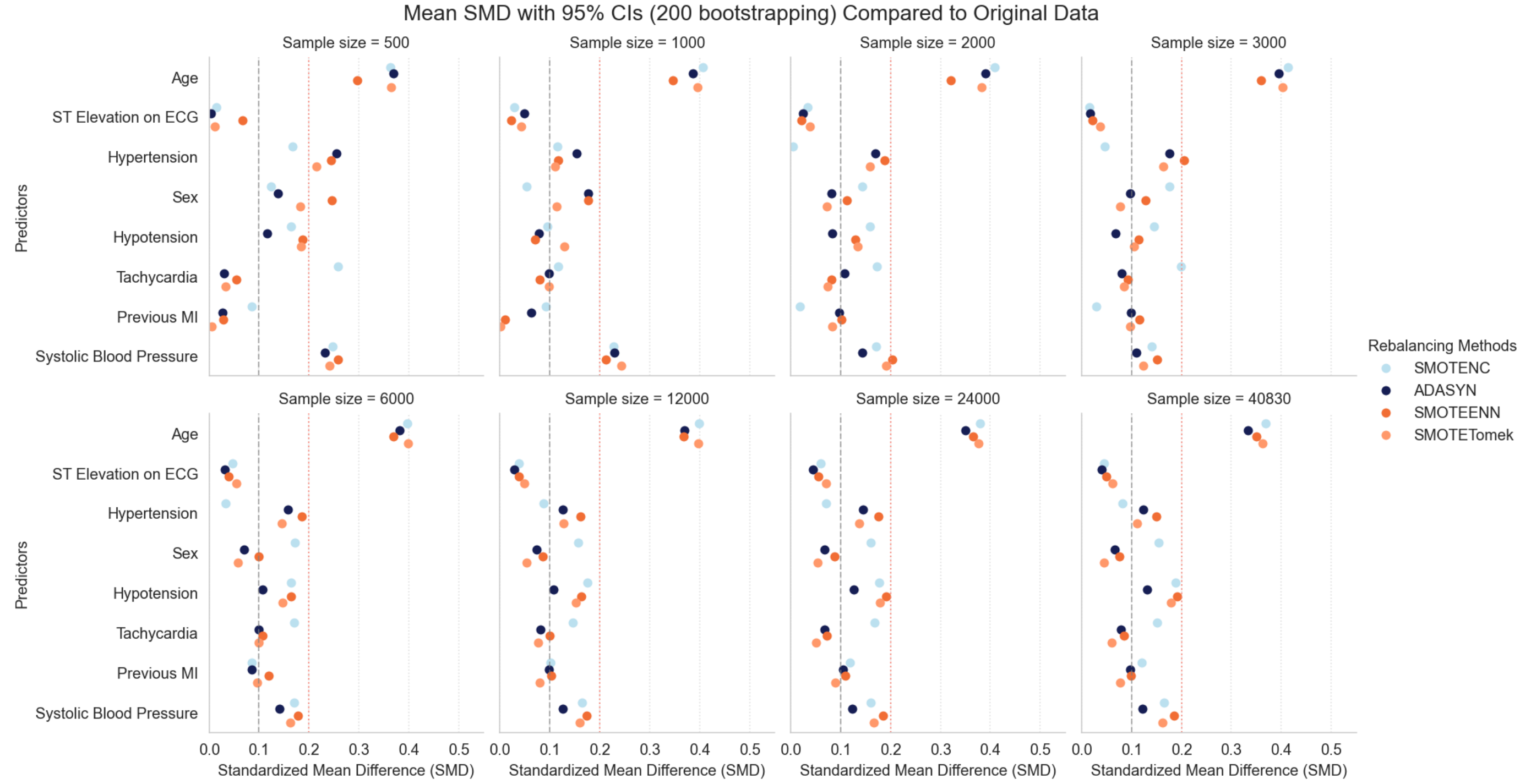

**Figure 2. Comparisons of the Original Imbalanced Data and Rebalanced Data.** The plots illustrate the mean SMDs with 95%CIs between the original data and 200 bootstrap resampling based on four data-level corrections: SMOTENC (light blue), ADASYN (dark blue), SMOTEENN (dark orange), and SMOTETomek (light orange). These corrections are applied to eight candidate predictors: age, ST elevation on ECG, hypertension, sex, hypotension, tachycardia, previous MI, and systolic blood pressure. Vertical dashed lines at SMD values of 0.1 and 0.2 indicate small and moderate differences, respectively. The extremely small variance observed across the 200 bootstrap iterations results in error bars for the 95% confidence intervals being too narrow to be visually distinguishable from the point estimates. **Abbreviations:** ADASYN, Adaptive Synthetic Sampling; ECG, Electrocardiogram; MI, Myocardial infarction; SMOTE, Synthetic Minority Over-sampling Technique; SMOTENC, SMOTE for Nominal and Continuous variables; SMOTENN, SMOTE and Edited Nearest Neighbors; SMOTETomek, SMOTE and Tomek links; SMD, Standardized Mean Difference.

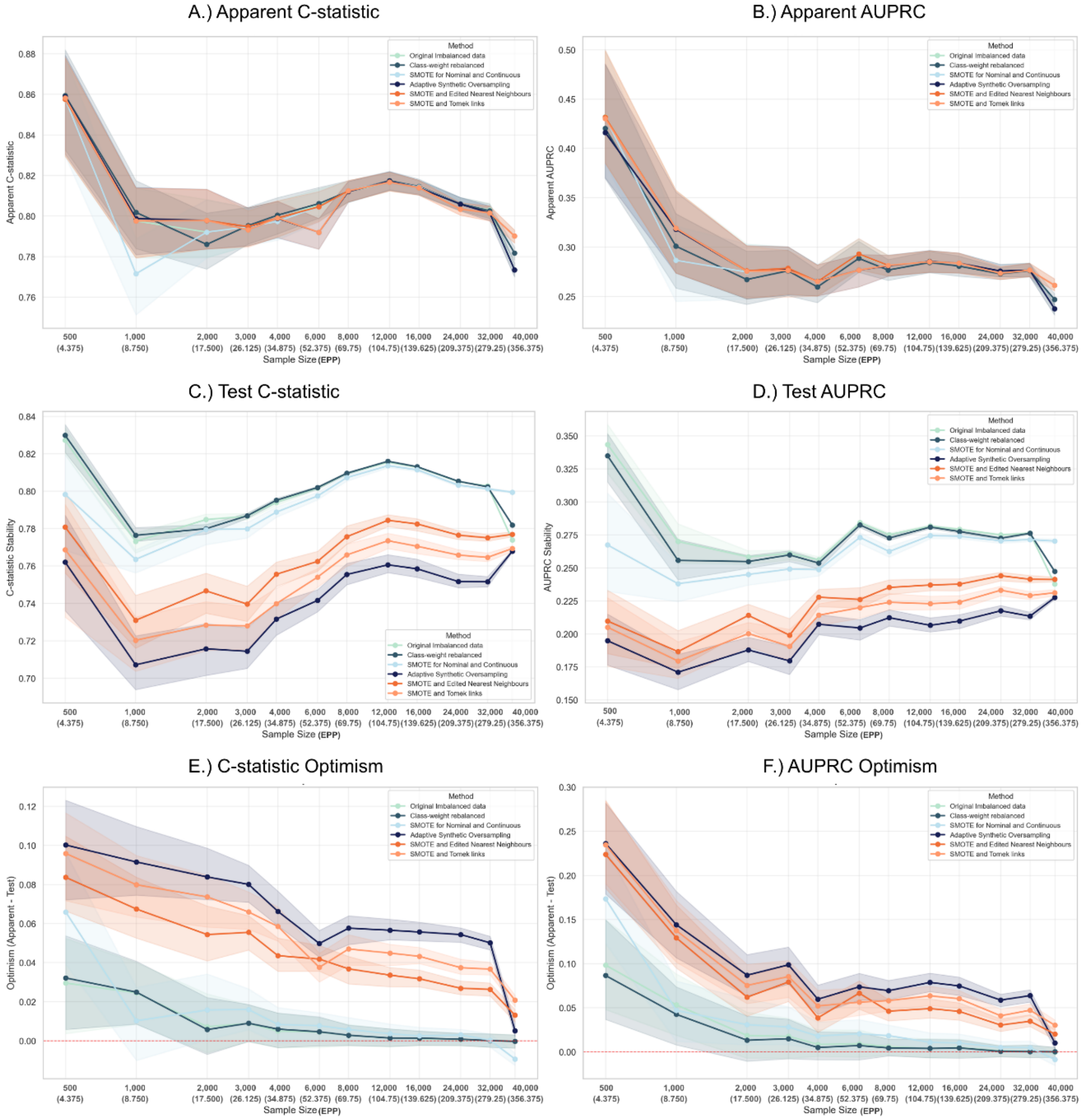

**Figure 3. Discrimination Performances (Median with IQR band) Across Sample Sizes and Methods.** The figure illustrates the (A) apparent C-statistic, (B) apparent AUPRC, (C) test C-statistic, (D) test AUPRC, (E) C-statistic optimism, and (F) AUPRC optimism for penalized logistic regression models. Performances were evaluated across simulated sample sizes ranging from 500 to 40,000, with the corresponding EPP displayed on the x-axis. Six methodological pathways for handling class imbalance were compared: original imbalanced data (baseline), class-weight rebalanced, oversampling via SMOTENC, oversampling via ADASYN, combined sampling via SMOTENN, and combined sampling via SMOTETomek. Apparent performance (Figures 3A and 3B) represented the training performance. Model internal validation and prediction stability (Figures 3C and 3D) were assessed by refitting the models using 200 BS datasets. Model optimism (Figures 3E and 3F) was determined by calculating the difference between the apparent and test performances. Shaded regions represent the variability across the bootstrap iterations. **Abbreviations:** ADASYN, Adaptive Synthetic Sampling; AUPRC, Area Under the Precision-Recall Curve; BS, Bootstrap; EPP, Events Per Predictor; SMOTE, Synthetic Minority Over-sampling Technique; SMOTENC, SMOTE for Nominal and Continuous variables; SMOTENN, SMOTE and Edited Nearest Neighbors; SMOTETomek, SMOTE and Tomek links.

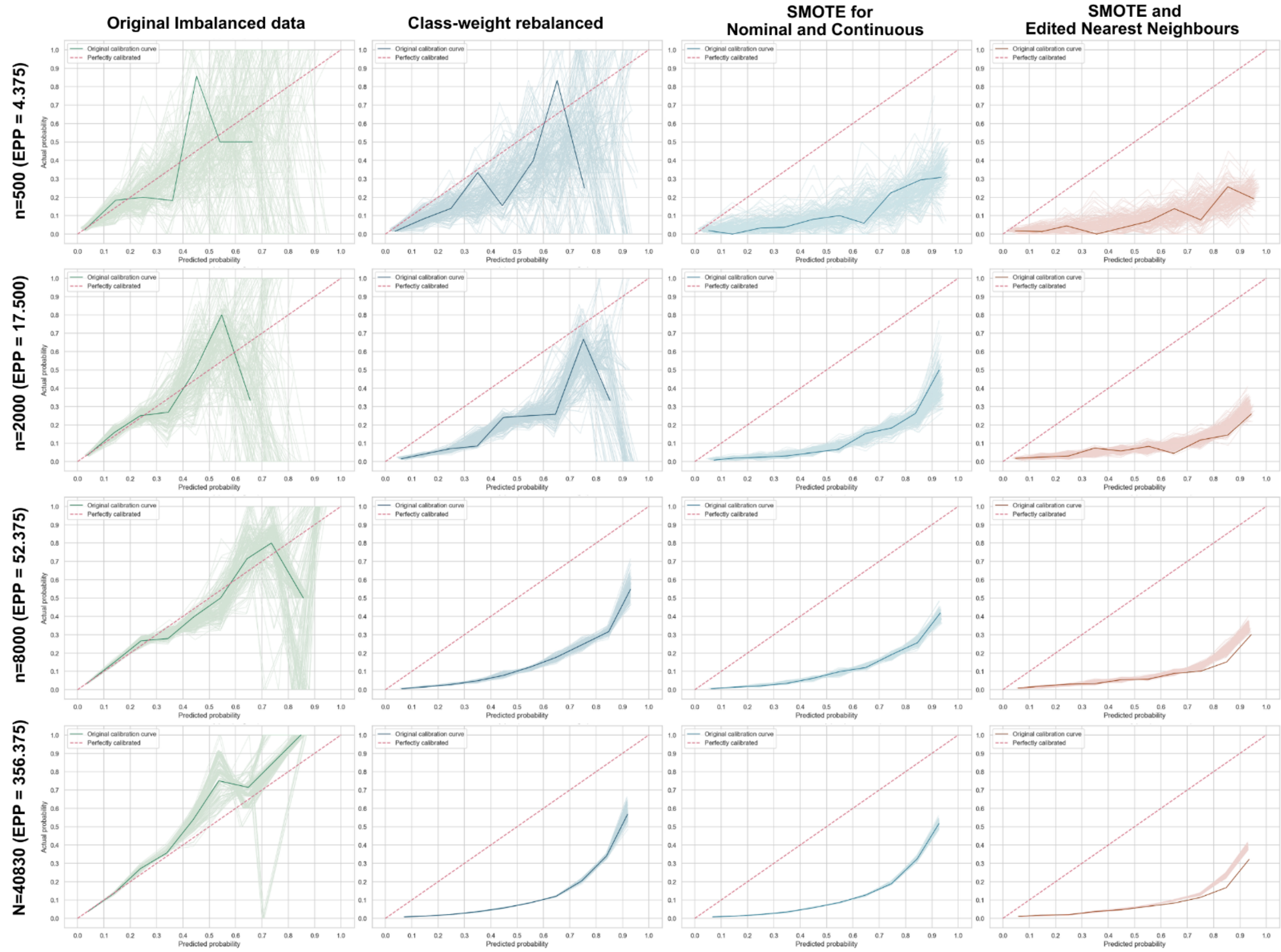

**Figure 4. Calibration Stability Plots Across Sample Sizes and Methods.** The calibration stability plots were visualized to compare model calibration, based on the predicted probabilities from 200 bootstrap resampling models evaluated using the original data. The figure displays four approaches to class imbalance, including the original imbalanced data as a baseline, class-weight rebalanced, SMOTENC (representative of oversampling approaches), and SMOTENN (representative of combined resampling approaches). Four landmark sample sizes are presented along with their respective EPP values. Within each plot, the solid bold line represents the original calibration curve; the lighter, transparent lines represent the variance across the 200 bootstrap iterations, demonstrating calibration stability; and the diagonal dashed red line represents perfect calibration. **Abbreviations:** EPP, Events Per Predictor; SMOTENC, Synthetic Minority Over-sampling Technique for Nominal and Continuous Variables: SMOTENN, Synthetic Minority Over-sampling Technique and Edited Nearest Neighbors.

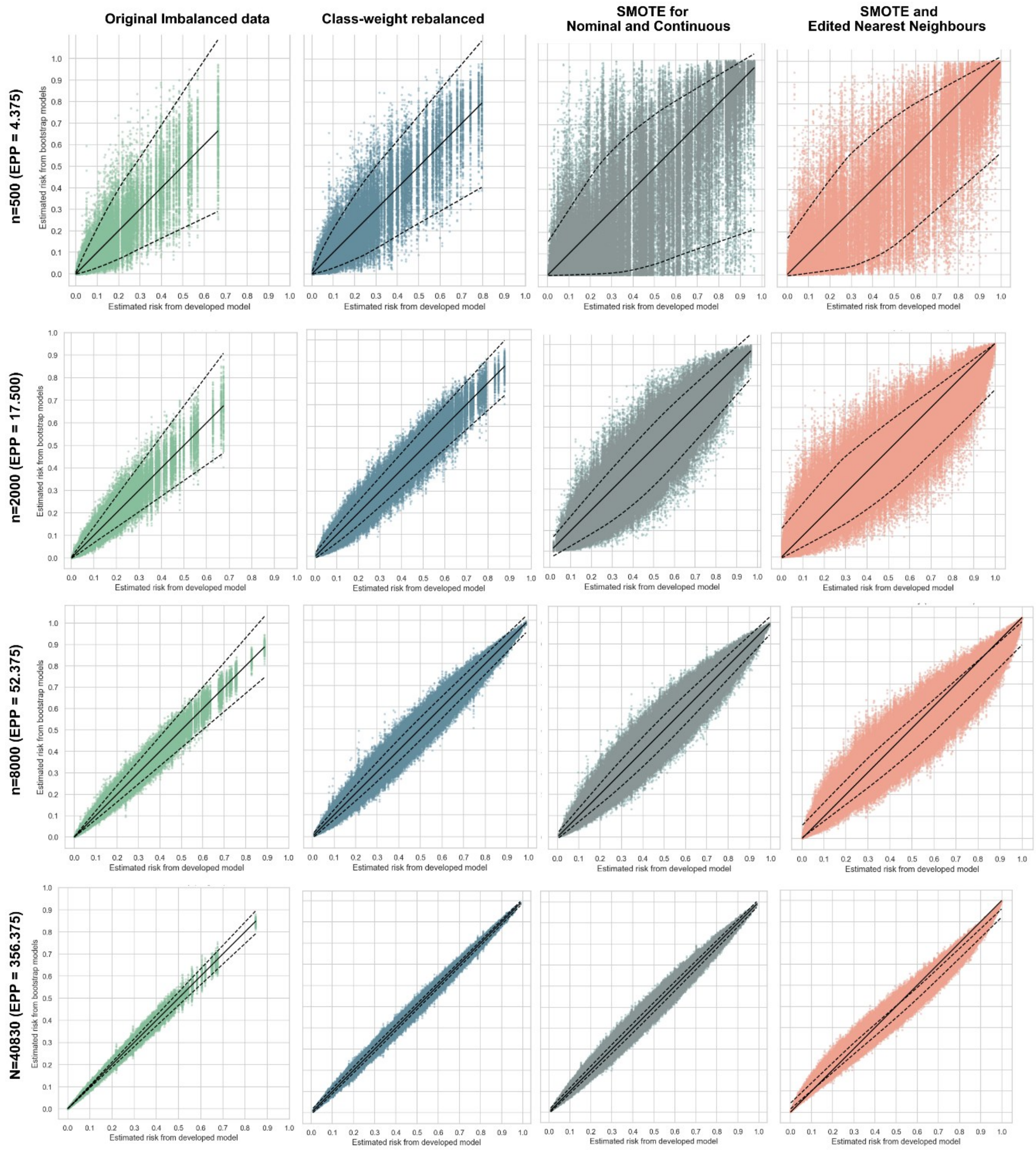

**Figure 5. Prediction Stability Across Sample Sizes and Method.** The prediction stability plots illustrate the variability of predicted probabilities based on the original data, comparing the developed models (on the X-axis) with 200 bootstrap resampling (on the Y-axis). The figure highlights four approaches to addressing class imbalance: the original imbalanced data as a baseline, class-weight rebalancing, SMOTENC (which represents oversampling techniques), and SMOTENN (which represents combined resampling techniques). Four landmark sample sizes are presented along with their respective EPP values. Within each plot, scatter points indicate the predicted probabilities across the bootstrap iterations; a solid bold line represents perfect agreement between the developed models and the bootstrap models; and the outer dashed lines show the 95% intervals of variability across the bootstrap iterations, illustrating prediction stability. **Abbreviations:** EPP, Events Per Predictor; SMOTENC, Synthetic Minority Over-sampling Technique for Nominal and Continuous Variables: SMOTENN, Synthetic Minority Over-sampling Technique and Edited Nearest Neighbors.

**Table S1.** Hyperparameter grid for a penalized logistic regression using GridsearchCV

| **Hyperparameter grids for 10-folds GridsearchCV** | |
|---|---|
| **Model:** LogisticRegression from *sklearn.linear_model* | |
| **General hyperparameters** | |
| Penalty methods | { 'l1', ‘l2’, ‘elasticnet’, None} |
| Solvers mapping with penalty methods | {'l1': 'saga',<br>'l2': 'lbfgs',<br>'elasticnet': 'saga',<br>None: 'lbfgs'} |
| Penalty strength (C) | [0.0001, 0.001, 0.01, 0.1, 1, 10, 100, 1000] |
| L1 ratio (for ‘elasticnet’) | [0, 0.25, 0.5, 0.75, 1] |
| Scoring | ‘roc_auc’ |
| Max iterations | 1000 |
| **Class weighting hyperparameters (for algorithm level correction)** | |
| Class weight | {'balanced',<br>[{0: 1, 1: v} for v in [2, 4, 6, 8, 10, 15, 20]} |

**Table S2.** Comparison of apparent c-statistics between no correction and imbalanced correction methods from model training on the original derived data and 200-boostrap sampling data across all sample size scenarios using a linear mixed model regression

| Variables | Adjusted mean differences (C-statistic: 0-100%) | 95%CIs | P-value |
|---|---|---|---|
| **Imbalanced Correction Methods** | | | |
| No correction | (Reference) | | |
| **Algorithm-level** | | | |
| Class weigthing | 0.257 | -0.041 to 0.556 | 0.091 |
| **Oversampling** | | | |
| SMOTENC | 0.004 | -0.294 to 0.303 | 0.978 |
| ADASYN | 0.092 | -0.206 to 0.391 | 0.545 |
| **Over- and under-sampling** | | | |
| SMOTENN | 0.004 | -0.294 to 0.303 | 0.978 |
| SMOTETomek | 0.027 | -0.272 to 0.325 | 0.861 |
| **Constant** | 85.450 | 85.210 to 85.690 | <0.001 |

Adjusted mean differences of c-statistics were obtained using a linear mixed model adjusted for full-interaction terms between imbalanced correction methods and sample sizes. **Abbreviations:** ADASYN, Adaptive Synthetic Oversampling; SMOTENC, Synthetic Minority Over-sampling Technique for Nominal and Continuous; SMOTENN, Synthetic Minority Over-sampling Technique and Edited Nearest Neighbours; SMOTETomek, Synthetic Minority Over-sampling Technique and Tomek-links.

**Table S3.** Comparison of test c-statistics between no correction and imbalanced correction methods from developmented models using 200-boostrap sampling and test on the original imbalanced data across all sample size scenarios using a linear mixed model regression

| Variables | Adjusted mean differences (C-statistic: 0-100%) | 95%CIs | P-value |
|---|---|---|---|
| **Imbalanced Correction Methods** | | | |
| No correction | (Reference) | | |
| **Algorithm-level** | | | |
| Class weigthing | 0.141 | -0.111 to 0.392 | 0.273 |
| **Oversampling** | | | |
| SMOTENC | -3.954 | -4.205 to -3.702 | <0.001 |
| ADASYN | -6.898 | -7.149 to -6.646 | <0.001 |
| **Over- and under-sampling** | | | |
| SMOTENN | -5.468 | -5.719 to -5.216 | <0.001 |
| SMOTETomek | -6.679 | -6.931 to -6.428 | <0.001 |
| **Constant** | 82.583 | 82.392 to 82.775 | <0.001 |

Adjusted mean differences of c-statistics were obtained using a linear mixed model adjusted for full-interaction terms between imbalanced correction methods and sample sizes. **Abbreviations:** ADASYN, Adaptive Synthetic Oversampling; SMOTENC, Synthetic Minority Over-sampling Technique for Nominal and Continuous; SMOTENN, Synthetic Minority Over-sampling Technique and Edited Nearest Neighbours; SMOTETomek, Synthetic Minority Over-sampling Technique and Tomek-links.

**Table S4.** Comparison of c-statistic optimism between no correction and imbalanced correction methods across all simulation scenarios using a linear mixed model regression

| Variables | Adjusted mean differences (Optimism of C-statistic: 0-100%) | 95%CIs | P-value |
|---|---|---|---|
| **Imbalanced Correction Methods** | | | |
| No correction | (Reference) | | |
| **Algorithm-level** | | | |
| Class weigthing | 0.116 | -0.193 to 0.425 | 0.460 |
| **Oversampling** | | | |
| SMOTENC | 3.957 | 3.649 to 4.267 | <0.001 |
| ADASYN | 6.990 | 6.681 to 7.299 | <0.001 |
| **Over- and under-sampling** | | | |
| SMOTENN | 5.472 | 5.163 to 5.781 | <0.001 |
| SMOTETomek | 6.707 | 6.397 to 7.015 | <0.001 |
| **Constant** | 2.866 | 2.629 to 3.104 | <0.001 |

Optimism was difference between apparent c-statistic (Model training performance on the original derived data) and test c-statistics (Test performance of each 200 boostrap models on the original imbalanced data). Adjusted mean differences of optimism were obtained using a linear mixed model adjusted for full-interaction terms between imbalanced correction methods and sample sizes. **Abbreviations:** ADASYN, Adaptive Synthetic Oversampling; SMOTENC, Synthetic Minority Over-sampling Technique for Nominal and Continuous; SMOTENN, Synthetic Minority Over-sampling Technique and Edited Nearest Neighbours; SMOTETomek, Synthetic Minority Over-sampling Technique and Tomek-links.

**Table S5.** Comparison of apparent AUPRC between no correction and imbalanced correction methods from model training on the original derived data and 200-boostrap sampling data across all sample size scenarios using a linear mixed model regression

| Variables | Adjusted mean differences (AUPRC: 0-100%) | 95%CIs | P-value |
|---|---|---|---|
| **Imbalanced Correction Methods** | | | |
| No correction | (Reference) | | |
| **Algorithm-level** | | | |
| Class weigthing | -1.479 | -2.135 to -0.823 | <0.001 |
| **Oversampling** | | | |
| SMOTENC | 0.030 | -0.626 to 0.687 | 0.927 |
| ADASYN | -1.411 | -2.068 to -0.755 | <0.001 |
| **Over- and under-sampling** | | | |
| SMOTENN | 0.030 | -0.626 to 0.687 | 0.927 |
| SMOTETomek | -0.108 | -0.764 to 0.687 | 0.746 |
| **Constant** | 44.194 | 43.683 to 44.705 | <0.001 |

Adjusted mean differences of AUPRCs were obtained using a linear mixed model adjusted for full-interaction terms between imbalanced correction methods and sample sizes. **Abbreviations:** AUPRC, the area under the precision and recall curve**;** ADASYN, Adaptive Synthetic Oversampling; SMOTENC, Synthetic Minority Over-sampling Technique for Nominal and Continuous; SMOTENN, Synthetic Minority Over-sampling Technique and Edited Nearest Neighbours; SMOTETomek, Synthetic Minority Over-sampling Technique and Tomek-links.

**Table S6.** Comparison of test AUPRC between no correction and imbalanced correction methods from developmented models using 200-boostrap sampling and test on the original imbalanced data across all sample size scenarios using a linear mixed model regression

| Variables | Adjusted mean differences (AUPRC: 0-100%) | 95%CIs | P-value |
|---|---|---|---|
| **Imbalanced Correction Methods** | | | |
| No correction | (Reference) | | |
| **Algorithm-level** | | | |
| Class weigthing | -0.760 | -1.026 to -0.494 | <0.001 |
| **Oversampling** | | | |
| SMOTENC | -7.491 | -7.756 to -7.225 | <0.001 |
| ADASYN | -14.551 | -14.816 to -14.285 | <0.001 |
| **Over- and under-sampling** | | | |
| SMOTENN | -13.148 | -13.414 to -12.883 | <0.001 |
| SMOTETomek | -13.857 | -14.122 to -13.591 | <0.001 |
| **Constant** | 33.992 | 33.788 to 34.197 | <0.001 |

Adjusted mean differences of c-statistics were obtained using a linear mixed model adjusted for full-interaction terms between imbalanced correction methods and sample sizes. **Abbreviations:** AUPRC, the area under the precision and recall curve;ADASYN, Adaptive Synthetic Oversampling; SMOTENC, Synthetic Minority Over-sampling Technique for Nominal and Continuous; SMOTENN, Synthetic Minority Over-sampling Technique and Edited Nearest Neighbours; SMOTETomek, Synthetic Minority Over-sampling Technique and Tomek-links.

**Table S7.** Comparison of AUPRC optimism between no correction and imbalanced correction methods across all simulation scenarios using a linear mixed model regression

| Variables | Adjusted mean differences (Optimism of AUPRC: 0-100%) | 95%CIs | P-value |
|---|---|---|---|
| **Imbalanced Correction Methods** | | | |
| No correction | (Reference) | | |
| **Algorithm-level** | | | |
| Class weigthing | -0.719 | -1.315 to -0.123 | 0.018 |
| **Oversampling** | | | |
| SMOTENC | 7.521 | 6.925 to 8.117 | <0.001 |
| ADASYN | 13.139 | 12.543 to 13.735 | <0.001 |
| **Over- and under-sampling** | | | |
| SMOTENN | 13.179 | 12.583 to 13.775 | <0.001 |
| SMOTETomek | 13.748 | 13.152 to 14.345 | <0.001 |
| **Constant** | 10.201 | 9.742 to 10.661 | <0.001 |

Optimism was difference between apparent c-statistic (Model training performance on the original derived data) and test c-statistics (Test performance of each 200 boostrap models on the original imbalanced data). Adjusted mean differences of optimism were obtained using a linear mixed model adjusted for full-interaction terms between imbalanced correction methods and sample sizes. **Abbreviations:** AUPRC, the area under the precision and recall curve**;** ADASYN, Adaptive Synthetic Oversampling; SMOTENC, Synthetic Minority Over-sampling Technique for Nominal and Continuous; SMOTENN, Synthetic Minority Over-sampling Technique and Edited Nearest Neighbours; SMOTETomek, Synthetic Minority Over-sampling Technique and Tomek-links.

**Combined Calibration/result Plots Across Sample Sizes and Methods**

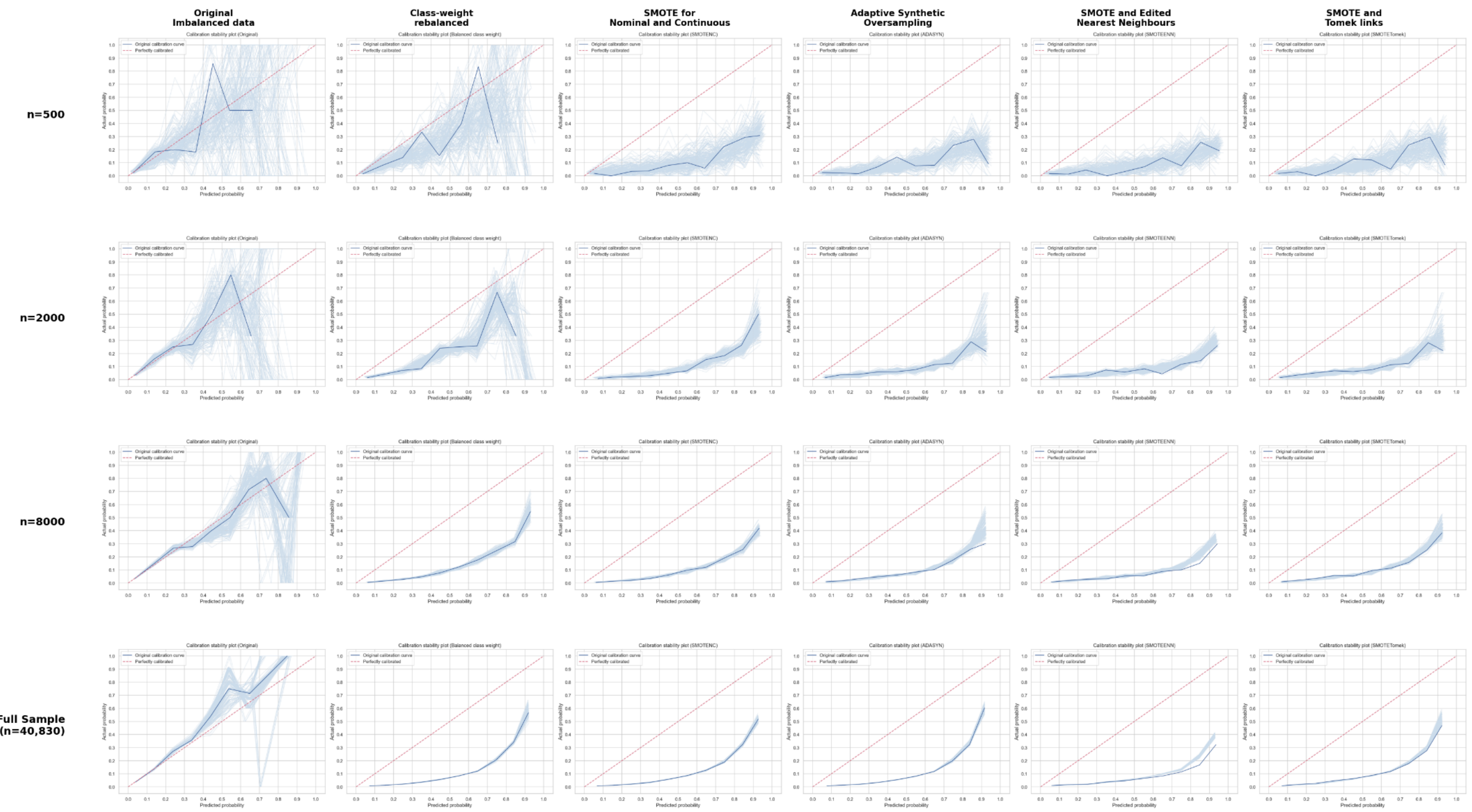

**Figure S1. Calibration Stability Plots Across Sample Sizes and All Methods.**

**Prediction Stability Across Sample Sizes and Methods**

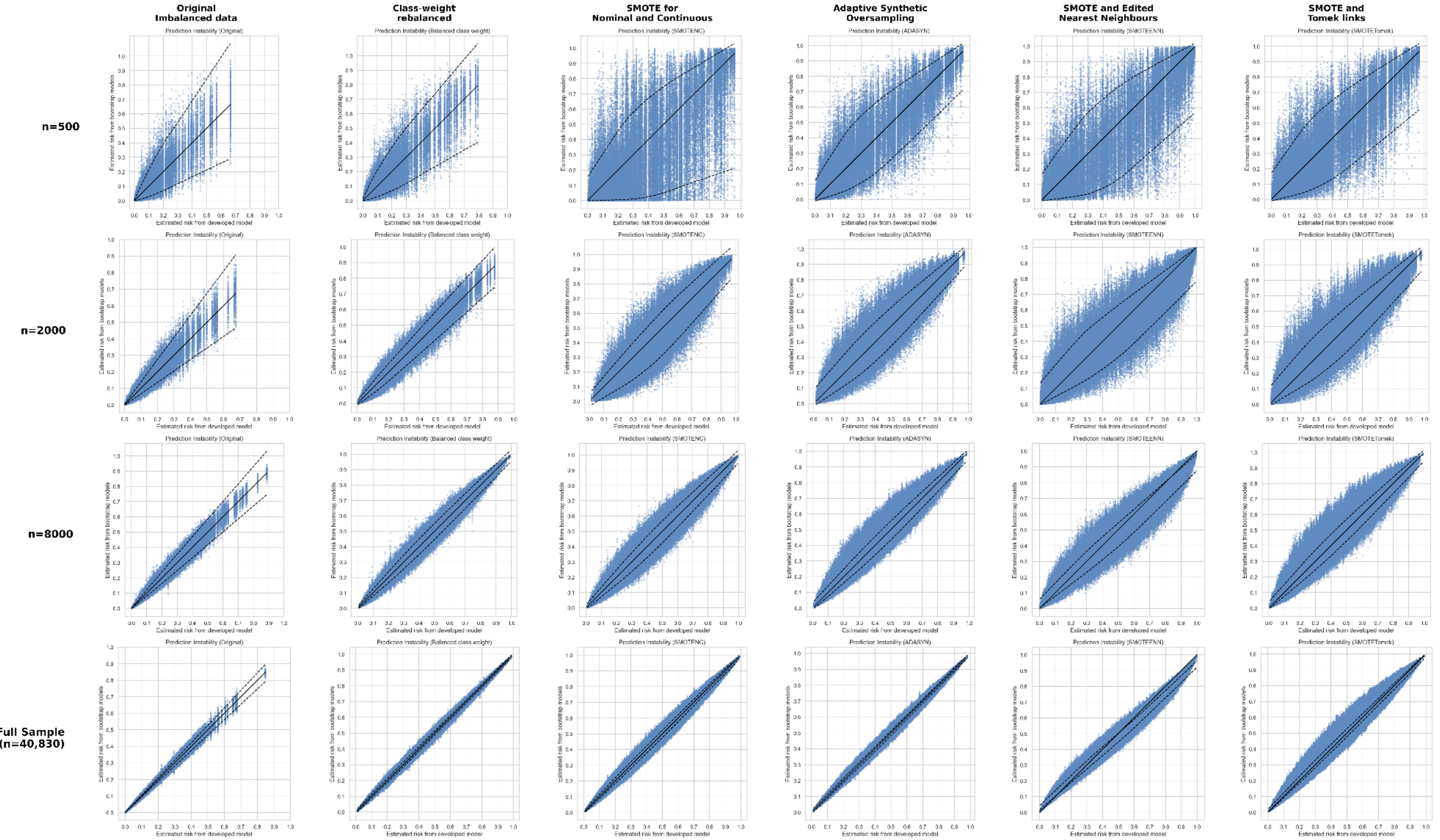

**Figure S2. Prediction Stability Across Sample Sizes and All Method.**

## MAPE Across Sample Sizes and Methods

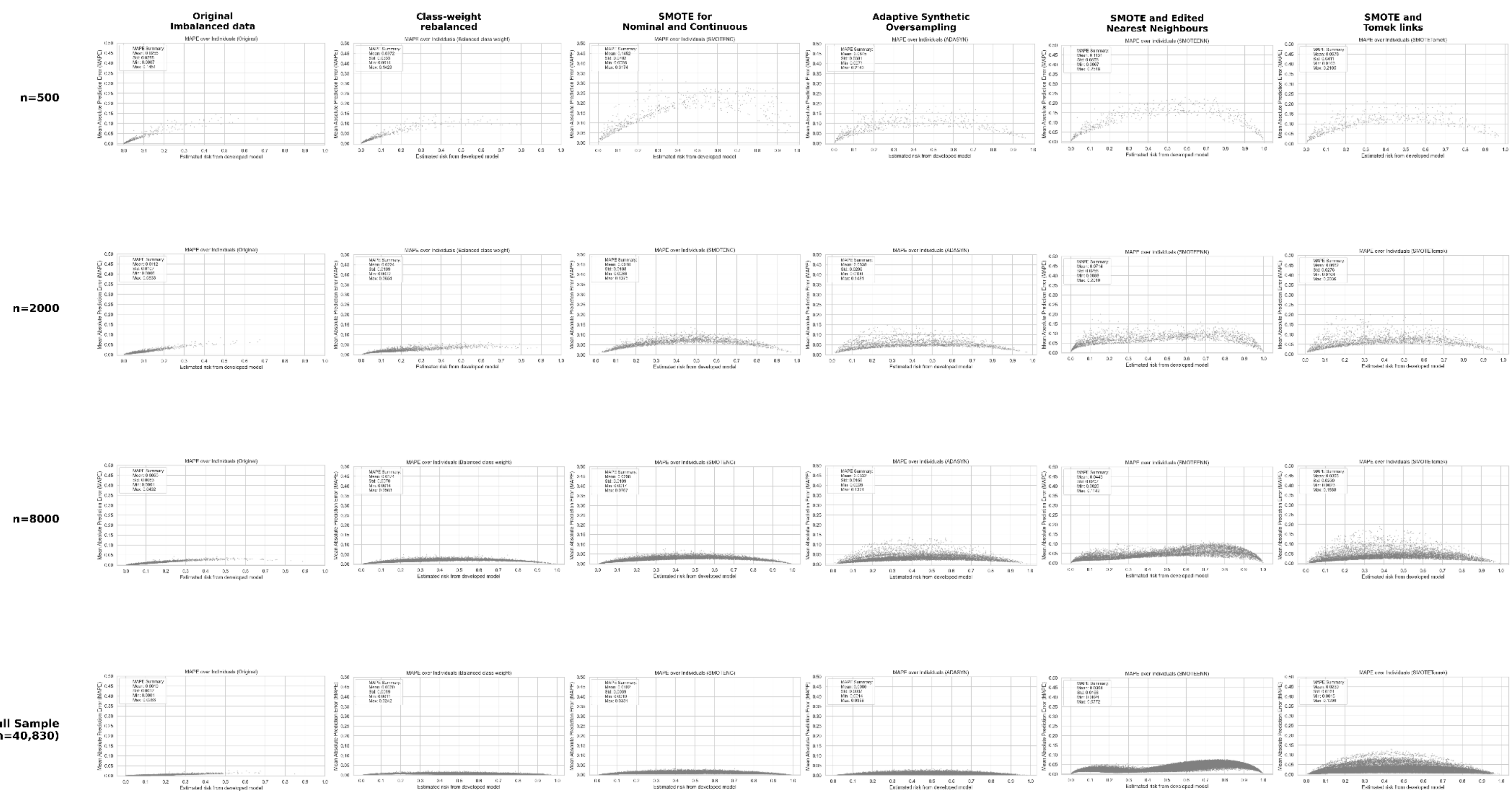

**Figure S3. Mean Absolute Prediction Error Across Sample Sizes and All Methods.**

**Classification Instability Index Across Sample Sizes and Methods**

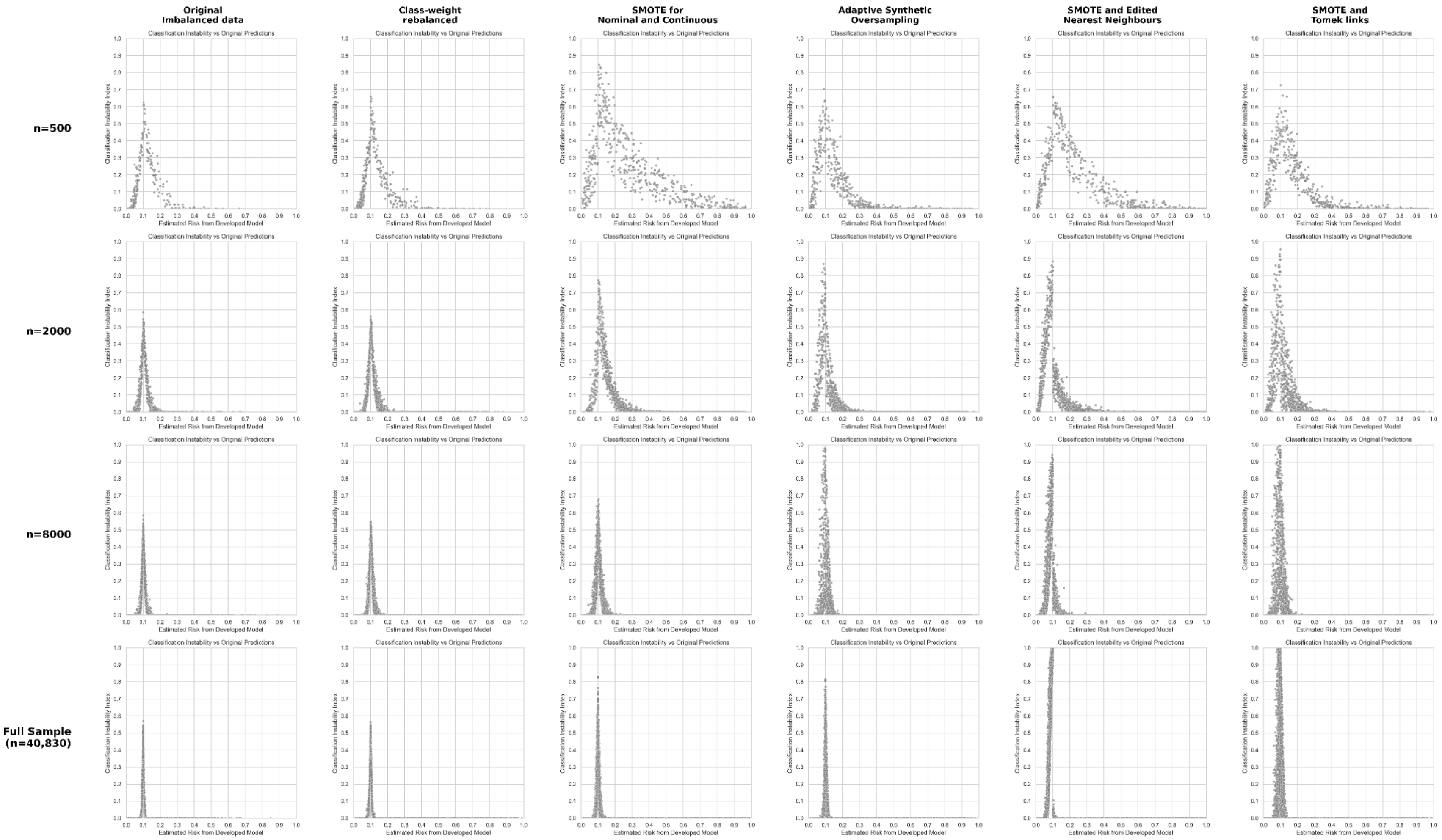

**Figure S4. Classification Instability Indices Across Sample Sizes and All Methods.**